\documentstyle[epsf]{l-aa}

\begin{document}

\thesaurus{}
\title{Nature of 60$\mu$m emission in 3C\,47, 3C\,207 and 3C\,334}
\author{Ilse M. van Bemmel\inst{1} \and Peter D. Barthel\inst{1} 
	\and Min S. Yun\inst{2}}
\offprints{bemmel@astro.rug.nl}
\institute{Kapteyn Astronomical Institute, P.O. Box 800, NL--9700~AV
	   Groningen, The Netherlands
	\and
	   NRAO VLA, P.O. Box 0, Socorro, NM, USA}
\date{Received date; accepted date}


\maketitle

\begin{abstract}

We try to explain the unusually high far-infrared emission seen by IRAS
in the double-lobed radio-loud quasars 3C\,47, 3C\,207 and 3C\,334. 
High resolution cm--mm observations were carried out to determine their
radio core spectra, which are subsequently extrapolated to the
far-infrared in order to determine the strength of the synchrotron
far-infrared emission.  The extrapolated flux densities being
considerably lower than the observed values, a significant nonthermal
far-infrared component is unlikely in the case of 3C\,47 and 3C\,334. 
However, this component could be responsible for the far-infrared
brightness of 3C\,207.  Our analysis demonstrates that nonthermal
emission cannot readily account for the difference between quasars and
radio galaxies in the amount of their far-infrared luminosity.  On the
other hand, a significant role for this mechanism is likely; full
sampling of the mm-submm spectral energy distributions is needed to 
address the issue quantitatively. 


\end{abstract}

\section{Introduction} \label{intro}

Thermal emission from cool circumnuclear dust is widely accepted as an
important mechanism for producing far-infrared radiation in radio-quiet
and radio-loud AGN (Sanders et al.  1986, Chini et al.  1989, Antonucci
et al.  1990).  This emission -- provided isotropic -- gives the
opportunity to test the orientation unification model of radio-loud
quasars (QSRs) and powerful radio galaxies (RGs) (Barthel 1989, Urry \&
Padovani 1995), since matched samples should yield similar far-infrared
output for both classes.  However, from IRAS observations it appears
that QSRs are stronger 60$\mu$m emitters than Fanaroff \& Riley class~II
RGs (Heckman et al 1992, 1994, Hes, Barthel \& Hoekstra 1995).  This
would imply that the simple unification model does not hold or that the
assumption of isotropic 60$\mu$m radiation is incorrect.  In support of
the latter, Pier \& Krolik (1992) drew attention to moderate anisotropy
effects in the thermal far-infrared radiation from optically thick tori
surrounding AGN.  Furthermore, Hoekstra, Barthel \& Hes (1997)
demonstrated that beamed nonthermal far-infrared radiation is not
insignificant in radio sources having prominent nuclei, and that the
60$\mu$m differences might be attributed to a stronger, beamed,
nonthermal component in the QSR class.  Within the unified model
framework, QSRs would -- at decreasing angle to the line of sight --
naturally lead into the radio-loud blazar class, objects in which the
dominance of the beamed component is without doubt.  Indeed, both
blazars and core-dominated QSRs display a smooth, single component
spectral energy distribution, extending from radio to X-rays, indicating
that all continuum radiation is of synchrotron origin (Landau et al. 
1986).  This causes blazars and core-dominated QSRs to be the most
luminous IRAS AGN (Impey \& Neugebauer 1988). 

IRAS detected three $z\sim0.5$, lobe-dominated 3CR QSRs at 60$\mu$m,
which corresponds to $\approx 40\mu$m emitted wavelength.  Comparable
radio galaxies were not detected -- contrary to the expectation within
the simple unification model.  The QSRs, 3C\,47, 3C\,207, and 3C\,334,
have relatively bright radio cores and large double-lobed structures. 
Two of these, 3C\,47 and 3C\,334, had been observed to display
superluminal motion -- a clear sign of beamed emission (e.g., Zensus \&
Pearson 1987) -- thus raising the suspicion that their far-infrared
brightness could be (partly) due to a beamed non-thermal component. 

The goal of the present research is to investigate to what extent
(beamed) nonthermal radiation can account for the infrared emission in
lobe-dominated QSRs.  We have determined the cm--mm core spectra of
3C\,47, 3C\,207, and 3C\,334 in order to assess their nonthermal
60$\mu$m radiation by means of extrapolation.  Given the possibility of
flaring submm components (observed in blazars -- e.g., Brown et al. 
1989) this research is a first attempt to isolate nonthermal and thermal
FIR radiation.  Observations over a wider frequency range are
forthcoming. 

\section{Sample description, observations and data reduction}

In order to determine the synchrotron far-infrared component in 3C\,47,
3C\,207, and 3C\,334, we observed their cm--mm core spectral energy
distribution, using the NRAO Very Large Array and the Owens Valley Radio
Observatory mm-array.  We supplemented these data with VLA Archive and
literature data, to examine core variability.  Some characteristics
\footnote{We use throughout H$_0 = 50$ km sec$^{-1}$ Mpc$^{-1}$, q$_0 =
0.5$} of the three quasars are listed in table 1.  The parameter $R_5$
specifies the fractional core flux density at 5~GHz emitted frequency. 
It should be noted that double-lobed 3C narrow-line radio galaxies of
similar radio power have typical $R_5$-values of order 0.01 (e.g.,
Fernini et al.  1997).  As mentioned above, 3C\,47 and 3C\,334 are
reported superluminal objects (Vermeulen et al.  1993, Hough et al. 
1992), while 3C\,207 is a suspected superluminal object (Hough, 1984). 
The last entry in table~1 specifies the measured superluminal component
speeds.  Images of the global QSR radio morphologies can be found in
Bridle et al.  (1994 -- 3C\,47 and 3C\,334) and Bogers et al.  (1994 --
3C\,207).  They all show large double-lobed structures and fairly
prominent one-sided jets. 

\begin{table}[t] 
\begin{center}
\begin{tabular}{lrrr}
\hline
\hline
{\bf source          }  &{\bf 3C\,47}   &{\bf 3C\,207 } &{\bf 3C\,334}\\
\hline
name (B1950)            &0133+207       &0838+133       &1618+177\\
$z$                     &0.425          & 0.684         & 0.555\\
$S_{60}$ (mJy)          &206            & 114           &126\\
log$P_{178}$ (W/Hz)	&28.36          &28.50          &28.21\\   
log$P_5^c$ (W/Hz) 	&25.68          &26.94          &26.10\\
$R_5$                   & 0.05          & 0.28          & 0.15(var.)\\
log$L_{60}$ (W)	 	&38.98          &39.16          &39.01\\
$v_{\rm SLM}$           & 7.4$c$        & suspected     & 3.2$c$\\
\hline
\end{tabular}
\caption[Basic properties of the quasars]{Basic properties of the
quasars.  $S_{60}$ is the IRAS flux density at 60$\mu$m, $P_{178}$ is
the total radio source power at 178~MHz (calculated adopting a spectral
index of 0.7), $P_5^c$ is the 5~GHz core radio power, $R_5$ is the
core/total flux density ratio at 5~GHz emitted frequency, calculated
adopting radio spectral indices of 0 and 0.7 for the core and extended
emission respectively.}
\end{center}
\end{table}

\subsection{VLA cm observations}

VLA A- and A/B-array observations at 4.9~GHz (C-band, 6~cm), 8.4~GHz (X,
3.6~cm), 15~GHz (U, 2~cm), 22~GHz (K, 1.2~cm) and 43~GHz (Q, 7~mm), were
conducted 1995 Sept.~5, 6, 7.  Two IFs were recorded at 50~MHz bandwidth
each.  Resolving VLA beams varied from 0.9 to 0.2~arcsec, at C- down to
K-band.  Q-band resolution was 0.4~arcsec, due the fact that only the
six inner telescopes of the array were equipped with Q-band receivers. 
High resolution observations are necessary in order to isolate core from
possible jet emission.  Single 3~minute snapshot scans were made at C-
and X-band, two scans of 3.5~minutes each at U- and K-band, and three
3.5~minute scans at Q-band.  Beam reference pointing at Q-band was
employed.  Secondary phase and amplitude calibrators were observed
before and after each scan. 

Primary calibrators were 3C\,84 and 3C\,286, with appropriate baseline
constraints, and inferred/adopted flux densities as listed in table~2. 
Both primary calibrators were observed during the run on 3C\,207 in
order to infer the absolute 3C\,84 flux density scale (3C\,84 was used
as primary calibrator for the 3C\,47 observations).  Calibration
uncertainty is dominated by the uncertainty in the absolute flux density
of the primary calibrators, which is $\approx 1\%$.  The array performed
well: antenna phase and amplitude calibration appeared stable to within
$\approx 0.1\%$, except for Q-band where these figures were $\approx
1\%$. 

\begin{table}[t]
\begin{center}
\begin{tabular}{lrr}
\hline
\hline
{\bf $\nu$}(GHz)&{\bf 3C\,84}	& {\bf 3C\,286} \\
\hline
4.9	& 27.63	& 7.426 \\
8.4	& 23.83	& 5.206	\\
15	& 23.63 & 3.502	\\
22	& 19.75	& 2.554	\\
43	& 12.47	& 1.472	\\
\hline
\end{tabular}
\caption{Input flux densities in Jy of the primary calibrators -- see text}
\end{center}
\end{table}

Data reduction was carried out in Groningen, using standard AIPS
routines.  The IF1 data being of inferior quality, analysis was
restricted to the IF2 data.  Tapering was applied for the (high
resolution) high frequency data, aiming for comparable resolution at all
bands.  All cores were detected, except for 3C\,47 in K- and Q-band, and
3C\,334 in Q-band.  For these we derived an upper limit, using the noise
levels.  The 3C\,47 K-band map was dominated by large phase errors and
therefore only a fairly high upper limit could be determined.  The noise
in all maps.  including the calibrators, was on the order of one to a
few mJy/beam, except for Q-band, where it was about 6~mJy/beam for
3C\,47 and 3C\,207.  Few Q-band antenna's were available for the 3C\,334
observations, yielding a low quality image and high Q-band flux density
upper limit.  We use the dirty map $5\sigma$ level as Q-band upper
limit.  Gaussian core profiles were fitted to determine the integral
flux densities, using the AIPS routine IMFIT.  The resulting values are
listed in table~3.  The ($3\sigma$) errors combine calibration and
fitting uncertainties. 

\begin{table}[t]
\begin{center}
\begin{tabular}{lrrr}
\hline
\hline
{\bf $\nu$}(GHz)& {\bf 3C\,47} 	&{\bf 3C\,207} 	&{\bf 3C\,334}	\\
\hline
4.9	& 77.9$\pm$3	& 532$\pm$21	& 159.5$\pm$6	\\
8.4	& 67.1$\pm$3	& 648$\pm$26	& 173$\pm$7 \\
15	& 50.4$\pm$2	&744$\pm$30	& 180$\pm$7 \\
22	& $\leq$42	&726.5$\pm$29	& 143.5$\pm$6 \\
43	& $\leq$30	&701.5$\pm$35	& $\leq$60	\\
\hline

\end{tabular}
\caption[Results]{VLA core flux density values in mJy from 
 Gaussian modelfits}
\end{center}
\end{table}

\subsection{OVRO 3mm observations}

Aperture synthesis observations of the 100~GHz continuum emission in
3C\,47, 3C\,207, and 3C\,334 were carried out with the Owens Valley
Millimeter Array on September 22, 1995, hence nearly simultaneously with
the VLA observations.  There were six 10.4 m diameter telescopes in the
array, each equipped with an SIS receiver cooled to 4K.  The receivers
were tuned to 100~GHz, and typical system temperatures of 300K (single
sideband) were achieved.  The array was in a compact configuration with
baselines of 15-65m E-W and 15-35m N-S, yielding a typical beam size of
14$''$(N-S) $\times$ 6$''$(E-W).  An analog correlator with 1~GHz total
bandwidth was used for these continuum mode observations.  Nearby
quasars were observed at 20 minute intervals to track the phase and
short term instrument gain, and Uranus ($T_B$=120K) was used for the
absolute flux calibration.  The data were calibrated using the standard
Owens Valley array program MMA (Scoville et al.  1993), while DIFMAP
(Shepherd et al.  1994) and AIPS were used for mapping and analysis. 
The uncertainty in the absolute flux measurement is about 15\%, and the
positional accuracy of the resulting images is better than $\sim 0.5''$. 

A 100~GHz nuclear continuum source is detected in all three quasars,
unresolved by the synthesized beam (see table~4).  The on-source
integration time was between 180 and 340 minutes, resulting in noise
sensitivity of 1-2 mJy/beam.  3C\,207 is a strong source at 100~GHz, and
the synthesized map is limited by dynamic range rather than by thermal
noise.  Evidence for lobe structures is seen in the visibility plots for
all three sources, but limited $uv$-coverage did not permit mapping of
these structures.  In 3C~47, two bright hot spots (features A and H in
Bridle et al.  1994) are clearly detected, located $\sim30''$ away from
the core.  However, their fluxes are uncertain because they lie near the
half-power point of the primary beam.  Contamination with arcsec scale
jet emission at 100~GHz is negligible, due to the spectral steepness of
jet radiation combined with the fact that the elongated synthesized beam
will not pick up substantial jet emission, as these are not oriented
N-S.

\begin{table} [t]
\begin{center}
\begin{tabular}{lr} 
\hline
\hline
{\bf source } & {\bf$S_{\rm 100GHz}$} \\
\hline
3C\,47           & $16.3\pm0.9$ \\
3C\,47 N.hotspot & $6.8^\dagger\pm0.9$ \\
3C\,47 S.hotspot & $15.4^\dagger\pm0.9$ \\
3C\,207          & $512\pm7$   \\
3C\,334          & $37.1\pm1.9$   \\
\hline
\end{tabular}   
\begin{list}{}{}
\item[$^\dagger$] 
 not corrected for primary beam attenuation.
\end{list}
\caption{OVRO 100~GHz core flux densities (mJy)}
\end{center}
\end{table}

\subsection{Archive and literature data}

Archival VLA data (1982--1983) were obtained from projects by Wardle,
Perley, and Ekers.  These data were reduced in the same way as our 1995
data.  To check for, and exclude, calibration induced variability, we
compared the archive short spacing flux densities with our 1995 values. 
Details on the observations and flux densities are listed in table~5. 

\begin{table}[t]
\begin{center}
\begin{tabular}{llllr}
\hline
\hline
{\bf source}	& {\bf observer}	&{\bf array}	&{\bf $\nu$}(GHz)	& {\bf $F$ } \\
\hline
3C47	& Ekers	 	& C	& 4.9	& 71$\pm$3	\\
3C207	& Wardle 	& A	& 4.9	& 477$\pm$19	\\	
3C207	& Wardle	& C	& 15	& 672$\pm$27	\\
3C334	& Perley	& C	& 4.9	& 165$\pm$7 	\\
3C334	& Perley	& C	& 15	& 99$\pm$4	\\
3C334	& Perkey	& C	& 22	& 54$\pm$3	\\
\hline
\end{tabular}
\caption{Arrays, bands and flux densities in mJy from VLA archive data -- see 
text for details.}
\end{center}
\end{table}

In addition to the archive VLA data, we examined literature data from
other telescopes, in search for variability during the epoch 1975--1995. 
These observations are listed in table~6.  Since for a number of
observations no precise value for the resolution is given, we listed the
telescopes used and whenever possible in which configuration. 

\begin{table}[t] \label{litdat}
\begin{center}
\begin{tabular}{lrrllr}
\hline
\hline
{\bf source}&{\bf $\nu$}(GHz)&{\bf $F$}&{\bf Ref.}&{\bf Resolution}&{\bf Year}\\
\hline
3C\,47    & 1.4   & 57.7  & 1	& VLA B+C & 85/86\\
3C\,47    & 4.8   & 72.8  & 1	& VLA A+B & 85/86\\
3C\,207	  & 1.4   & 258   & 9   & VLA A   & 92  \\
3C\,207   & 4.8   & 395   & 4	& VLA	& 80	\\
3C\,207   & 5     & 510   & 2	& MRAO$^a$ & 74$^b$\\
3C\,207   & 5     & 490   & 3	& transatl.VLBI	& 81 \\
3C\,207   & 10.6  & 570   & 4	& VLA	& 80	\\
3C\,207	  & 10.6  & 588   & 10   & VLA    & 86$^b$  \\	
3C\,334   & 1.4   & 134   & 8    &  VLA A  & 80/81  \\
3C\,334   & 4.9   & 138   & 6    & VLA A+B & 86    \\
3C\,334   & 5     & 170   & 7    &  MRAO & 77$^b$  \\
3C\,334   & 5     & 110   & 3    & transatl.VLBI & 81    \\
3C\,334   & 10.6  & 83    & 5    & transatl.VLBI & 83    \\
3C\,334   & 10.6  & 96    & 5    & transatl.VLBI & 88    \\
\hline   
\end{tabular}
\end{center}

\vspace{1mm}
{\footnotesize $^a$ MRAO (Cambridge, UK) 5~km telescope \\
 $^b$ date of publication}

\caption[Literature data]{Literature references, with year of observation. 
All fluxes are in mJy; column~4 gives the literature reference.
References: 1=Fernini et al. (1991); 2=Pooley \& Henbest (1974);
3=Barthel et al. (1984); 4=Rudnick et al. (1986); 5=Hough et al. (1992); 
6=Bridle et al. (1994); 7=Jenkins et al. (1977); 8=Hintzen et al. (1983);
9=Bogers et al. (1994); 10=Hough (1986).}
\end{table}

\section{Results and spectral fits} 

\subsection{Core cm--mm spectra}

The core spectra for 3C\,47, 3C\,207 and 3C\,334 are plotted in
figure~1, using data from tables 3, 4, 5, and 6.  Our 1995 data, which
will be used for the spectral energy distribution (SED) fits, are
plotted as open circles.  The unusually high IRAS 60$\mu$m points appear
on the right in each figure as filled circles.  Literature data are
plotted as stars, the 1982/1983 VLA archive data as filled triangles. 
All errors are within the sizes of the points.  The following remarks
can be made for the individual objects. 

\begin{figure*}[t] 
\begin{center}
\centering
\centerline{\epsfysize=6truecm
	\epsfbox[78 514 542 688]{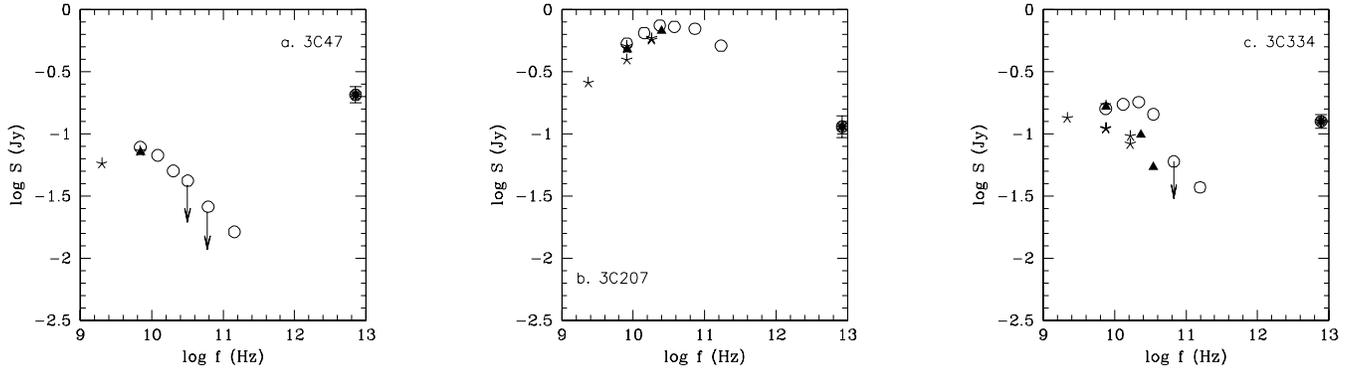}
	\message{Figuur 1}}
\caption{Rest frame core spectra for the IRAS detected quasars,
including literature and 1983 VLA data.  Open circles represent our 1995
data, filled triangles 1983 VLA data, and stars literature data.  The
filled circles represent the total (thermal plus nonthermal) IRAS
$60\mu$m flux densities.  All error bars fall within the size of the
points, except for the IRAS observations.}
\end{center}
\end{figure*}

\subsubsection{3C\,47}

An archival VLA 5~GHz dataset (observed by Ekers in June 1982, C
array) appeared in good agreement with our observations: no evidence for
core variability is seen.  Also the literature data show no sign of cm
variability over the last twenty years.  Since this quasar seems to have
a stable core, we decided to include the 1.4~GHz datapoint (1985
observation, VLA A-array) from Fernini et al.  (1991) in our fitting
procedure. 

\subsubsection{3C\,207}

For this source we retrieved VLA archive observations at C- and U-band,
made by Wardle in March 1982 (A-array) and May 1983 (C-array)
respectively.  The core flux densities are in good agreement with our
1995 VLA data.  Also, most literature data show no large discrepancies
with respect to our data.  However, Hough (1986) reports changes of
$\sim$ 100 -- 200 mJy in the early 1980's.  Comparison of transatlantic
VLBI with VLA data at 5~GHz is also indicative of some level of
variability in the 3C\,207 nucleus.  Since the core shows no strong
variability, we added the (1992 VLA A-array) 20~cm flux density value
from Bogers et al.  (1994). 

\subsubsection{3C\,334}

An extensive archival VLA data set was available for this quasar,
observed by Perley in January 1983 (C-array).  Observations were carried
out in C-, U-, and K-band.  3C\,334 displays substantial core
variability with respect to our 1995 data, as seen from figures 1 and 3. 
Figure~3 displays 5~GHz core flux density values, measured at $\sim
0.5''$ resolution over the last two decades (data from table~2 in Hough
et al.  1992). 

\subsection{Core spectral energy distributions}

\begin{figure*}[t]
\begin{center}
\centering
\centerline{\epsfysize=6truecm
	\epsfbox[78 514 542 688]{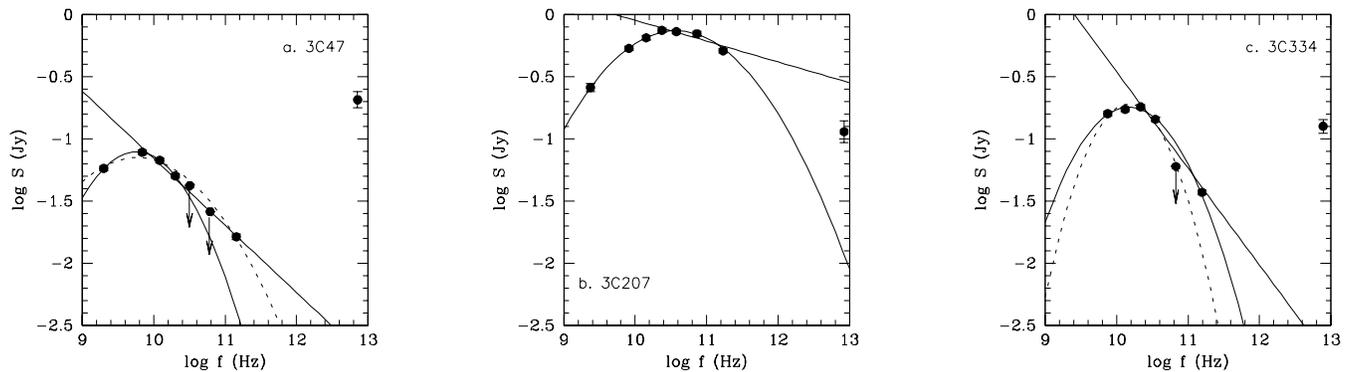}
	\message{Figuur 2}}
\caption{Rest frame fits for the IRAS detected quasars, using only our
1995 data and additional 20cm points for 3C\,47 and 3C\,207.  The
straight lines represent powerlaw fits to the high frequency points, the
parabola fits draw from an empirical result for blazar spectra.  The
dashed parabola in figure~a includes the 3mm point, while the solid
parabola does not.  See text for details on the fitting procedure.}
\end{center}
\end{figure*}

To obtain a good estimate of the (beamed) nonthermal 60$\mu$m flux
density, we fitted curves to the nearly simultaneous 1995 cm and mm
data.  As mentioned above, for 3C\,47 and 3C\,207 we added the 20cm
(L-band) points from Fernini et al.  (1991) and Bogers et al.  (1994)
respectively.  Upper limits are excluded from the fitting procedures. 
The fitted curves can subsequently be extrapolated into the
far-infrared.  However, the shape of the curve is a priori unknown.  We
therefore used two different models.  First we fitted a parabola, which
is an empirical result for blazar spectra (Landau et al., 1986).  In the
light of the unification theory and provided we are dealing with single
components, these parabolic fits can be used to fit core spectra for
double-lobed quasars.  It should be noted however that Brown et al. 
(1989) found evidence for double components in their blazar sample.  We
will come back to this issue in section 4.1. 

Second we used the theoretical spectrum of a single self-absorbed
synchrotron source, which has a $F \sim \nu^{\alpha}$ powerlaw at
frequencies larger than the turnover frequency.  For this fit we
excluded all the frequencies below the turnover frequency.  Both fits
were done using least-squares methods.  The resulting fits, in the QSR
rest frames, are shown in figure~2.  With the exception of 3C\,47, the
parabolae fit remarkably well, but it should be kept in mind that their
high frequency parts are not well constrained.  

Given the upper limits at 1cm and 7mm, a parabola does not yield a good
fit to the 3C\,47 data when we include the 3mm point (dashed parabola in
figure~2a).  Leaving out this 3mm point, a single parabola does fit the
3C\,47 data remarkably well (solid parabola in figure~2a).  A secondary
(sub)mm component may be causing the rather high 3mm point.  In the case
of 3C\,334, the parabola does fit well (solid parabola in figure~2c),
but the Q-band upper limit is below our fit.  If we exclude the 3mm
point from the fit procedure, the parabola does not change
significantly.  The dashed parabola is a fit including the Q-band upper
limit and excluding the 3mm point. 

As can be seen immediately, only in 3C\,207 the IRAS flux density may
suffer from substantial nonthermal contamination, adopting the high
frequency powerlaw fit.  For the other two quasars the far-infrared
point is far above the extrapolated nonthermal value, leaving a factor
of 100 to account for.  Note that especially 3C\,47 stands out: the IRAS
point is even higher than the radio peak flux density.  The extrapolated
values of the FIR flux density for different models are compared with
the observed values in table~7.  $S_{60}$ is the IRAS flux density from
Hes et al.  (1995), $F_e$ are the extrapolated flux densities and
$\Delta F_{60}$ is the FIR excess defined as the difference between
observations and models at 60$\mu$m observed wavelength.  All values are
in mJy/beam. 

\begin{figure}[t] 
\begin{center}
\leavevmode
\hbox{%
\epsfysize=9cm
\epsfbox{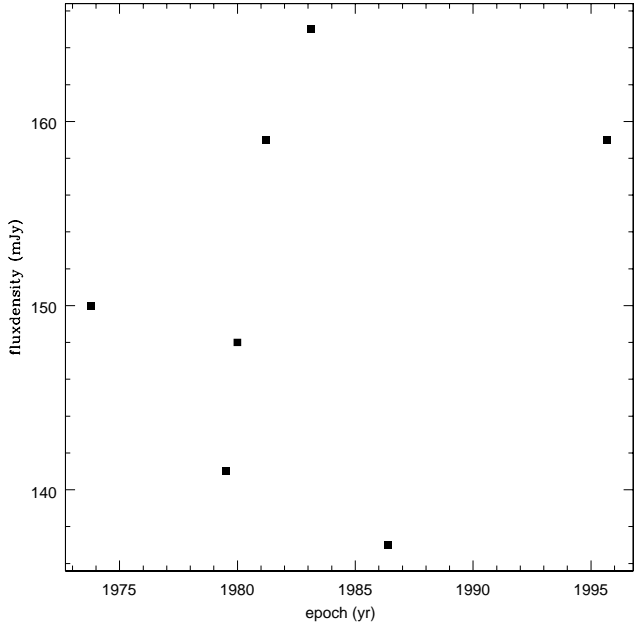}}
\message{Figuur 3}
\caption{Variability of 3C\,334 over the last two decades. Data points
taken from table~2 in Hough et al. (1992)}
\end{center}
\end{figure}

\begin{table}[t] 
\begin{center}
\begin{tabular}[H]{lrrrr}
\hline
\hline
{\bf source} & {\bf $S_{60}$} & {\bf $F_{e,par}$}$^a$ & {\bf $F_{e,pl}$}$^b$ 
 &{\bf$\Delta F_{60}$}\\
\hline
3C\,47  & 206$\pm$33   & $\sim 0$         & 2.0$\pm$0.1   & $\sim$204     \\
3C\,207 & 114$\pm$26   & 11.5$\pm$0.1	  & 291$\pm$5     & $\geq$--177 \\
3C\,334 & 126$\pm$17   & $\sim 0$         & 1.9$\pm$0.1   & $\sim$124 \\
\hline
\end{tabular}

\vspace{1mm}

{\scriptsize $^a$ extrapolated flux density from parabola fit}\\
{\scriptsize $^b$ extrapolated flux density from powerlaw fit}\\
\caption[Comparison of FIR flux densities]{
Comparison of measured total and extrapolated nonthermal
FIR flux densities, in mJy.}
\end{center}
\end{table} 

\section{Analysis and discussion} \label{analysis}

Since there is a substantial time interval between the (1983) IRAS
observations and the other data, core variability could be a possible
explanation for the FIR excess.  However, long-term radio data from
literature show that the core variations do not exceed $\approx$10\% and
therefore are not large enough to explain the total excess.  In
addition, all variability references report fairly low core flux
densities for 1983 (IRAS epoch), which does not help to explain a FIR
{\em excess}.  Although the IRAS detections could be spurious, since the
sources are close to the detection limit, or due to a confusing source,
the issue remains why no spurious radio galaxy detections appear.  Below
we examine all possible explanations for the FIR excess observed in
these quasars. 

\subsection{Non-thermal FIR}

The main goal of this research was to find out if beamed non-thermal
radiation could be responsible for the FIR excess.  Adopting single
component core models it is evident from the plots in figure~2 that
beaming cannot be responsible, except for 3C\,207 applying the powerlaw
fit.  However, we do know that beaming operates in 3C\,47 and 3C\,334,
both superluminal objects.  Both the FIR beaming and the radio core
R-parameter depend on the Doppler factor $\Gamma$ to some power.  In
terms of R-parameter strength the strongest component of nonthermal FIR
radiation would be expected from 3C\,207, followed by 3C\,334 and
3C\,47.  This order is not reflected in the total 60$\mu$m luminosities:
the largest FIR excess, normalized with respect to the 5~GHz flux
density, originates in the quasar with the smallest core fraction. 
While the reported superluminal velocities roughly scale with the
infrared excess in 3C\,334 and 3C\,47, a measurement is still lacking in
3C\,207.  Moreover, these tests have little statistical significance. 

The formula derived by Hoekstra et al.  (1997) permits an estimate of
the relative amount of beamed radiation.  Hoekstra et al.  use a
relation depending on $Q$ and $C_{60}$ to estimate the amount of beamed
60$\mu$m emission.  Here $Q$ is the observed 5GHz core fraction, and
$C_{60}$ is a measure for the value of $Q$ at which beaming becomes
significant.  They determined the value of $C_{60}$ from a large sample
of FIR-detected blazars and quasars, including the three discussed here. 
This yields a maximum nonthermal FIR contribution of 15\%, for 3C\,207,
and considerably less for 3C\,47 and 3C\,334. 

The Hoekstra et al.  (1997) analysis postulates a direct (single
component) connection between the nonthermal radio and FIR emission.  It
is quite likely that the real situation is more complicated.  As
mentioned above, Brown et al.  (1989) demonstrated the presence of two
nonthermal core components in blazars.  One component is fairly
quiescent and dominates the radio to mm region. A second, ultracompact, 
component is prevalent in the submm regime. This component becomes 
self-absorbed at wavelengths longer than $\sim$ 3mm, and displays strong 
variability (flares), sometimes within days. Our data do not constrain the submm
region well: it is likely that such variable submm components are being
missed in our analysis.  The possibility that such a component has
pushed the total (thermal plus nonthermal) FIR into IRAS detection
cannot as yet be ruled out.  In fact, as inferred from the 100~GHz data,
our 3C\,47 observations may have caught such a submm component. In 3C\,334 
this might also be the case, considering that our Q-band upper limit is
lower than the predicted value if we include the 3mm point in our fit.

We stress that since the IRAS detections are just above the detection
limit, not the total 60$\mu$m emission has to be accounted for, but only
a substantial nonthermal component lifting the total 60$\mu$m flux into
IRAS detection.  For 3C\,47 and 3C\,334, superluminal and hence beamed
objects, the possibility of an additional, variable submm component is
considered likely.  Full sampling of the cm-mm-submm-FIR spectral range
is needed to confirm our suspicion. 

\subsection{Thermal FIR}

If most or all 60$\mu$m emission is identified with thermal radiation,
the question arises why 3C\,47, 3C\,207 and 3C\,334 are more luminous
than other quasars and radio galaxies.  There are several mechanisms
that can produce thermal FIR in AGN: cold cirrus and warm starburst
heated dust in the host galaxies, and furthermore warm AGN-related dust. 
Models for these components are for instance described by Rowan-Robinson
\& Crawford (1989).  In order to distinguish between these, more IR data
are needed to perform a detailed analysis of the FIR spectral energy
distribution.  Multiple component fitting is necessary to isolate the
various dust components.  However, if these mechanisms are responsible
for the FIR excess, unification is difficult, since it states that radio
galaxies and lobe-dominated quasars are basically the same objects and
thus should have similar dust composition. 

If we wish to maintain the unification concept we have to postulate
optically thick dust emission at $\sim40\mu$m in combination with aspect
effects.  If the FIR excess originates from an optically thick torus
shielding the optical QSO in a plane perpendicular to the radio axis,
models by Pier \& Krolik (1992) yield aspect dependent anisotropies of
factors up to about ten.  At longer FIR wavelengths the anisotropies
should become zero (optical thin dust): data at 100$\mu$m or longer
wavelengths are needed to investigate optical thickness effects. 

\section{Conclusions and future research} \label{conclusions}

We cannot readily explain the extraordinary high FIR flux density
observed in the 3C\,47, 3C\,207, and 3C\,334.  While relativistic
beaming of a single nonthermal component cannot account for the total
FIR radiation, the radiation of relativistically beamed multiple core
components is likely to contribute significantly in the far-infrared. 
Thermal mechanisms could play a role, in combination with optical
thickness and aspect effects, but we need more data in the submm and
infrared to address these in detail.  We are currently engaged in
measurements of matched pairs of radio galaxies and quasars, comparing
their thermal FIR output by combining cm, mm, submm and FIR data from
VLA, JCMT, and ISO observations.  In the meantime, we have little doubt
that the importance of nonthermal FIR and submm radiation in double-lobed 
radio sources has been underestimated. 

\bigskip
\noindent

{\bf Acknowledgments}
\noindent
We acknowledge initial involvement in this work of R.~Hes and
H.~Hoekstra, OVRO support by N.~Scoville and A.~Sargent, and the use of
archival VLA data from R.~Ekers, R.~Perley, and J.~Wardle.  The NRAO VLA
is operated by Associated Universities, Inc.  under contract with the
National Science Foundation. OVRO research is supported in part by NSF 
Grant AST~93-14079.

\end{document}